\documentclass[pra,a4paper,twocolumn,showpacs,groupedaddress]{revtex4-1}
\usepackage{graphicx}
\usepackage{amsmath}
\usepackage{epstopdf}

\begin{document}

\title{Control of populations of two-level systems by a single resonant laser pulse}

\author{Nikolay V. \surname{Golubev}}
%\email[e-mail: ]{nikolay.golubev@pci.uni-heidelberg.de}
\affiliation{Theoretische Chemie, Universit\"at Heidelberg, Im Neuenheimer Feld 229, 69120 Heidelberg, Germany}

\author{Alexander I. \surname{Kuleff}}
\email[e-mail: ]{alexander.kuleff@pci.uni-heidelberg.de}
\affiliation{Theoretische Chemie, Universit\"at Heidelberg, Im Neuenheimer Feld 229, 69120 Heidelberg, Germany}

\date{\today}

\begin{abstract}
We present a simple approach allowing to obtain analytical expressions for laser pulses that can drive a two-level system in an arbitrarily chosen way. The proposed scheme relates every desired population-evolution path to a single resonant laser pulse. It allows to drive the system from any initial superposition of the two states to a final state having the desired distribution of the populations. We exemplify the scheme with a concrete example, where the system is driven from a non-stationary superposition of states to one of its eigenstates. We argue that the proposed approach may have interesting applications for designing pulses that can control ultrafast charge-migration processes in molecules. Although focused on laser driven population control, the results obtained are general and could be applied for designing other types of control fields.
\end{abstract}


\maketitle

%\section{Introduction}

Two-level systems, quantum systems possessing only two eigenstates, have played an important role in quantum mechanics since its advent. Although rarely existing in nature in their pure form, they often serve as models in many areas of physics and are used successfully for describing a large variety of physical phenomena. A special class of problems are the driven two-level systems, i.e. situations when an external field is applied with the aim to control the quantum evolution of the system. Such situations can be encountered in nuclear magnetic resonance techniques \cite{NMR_rev_RMP05}, Josephson-junction circuits \cite{Nakamura_PRL01}, spin rotations in quantum dots \cite{Greilich_etal_NatPhys09} and qubit control in general, as well as in laser-induced population transfer in atoms and molecules \cite{Vitanov_ARPC01,Shapiro_RMP07}. 

At the same time, designing laser fields for reaching a selected state of a quantum system has attracted a lot of efforts in the last few decades. Different schemes for control of the quantum dynamics were proposed, like $\pi$- and chirped pulses \cite{pipulse_JCP00}, stimulated Raman adiabatic passage (STIRAP) \cite{Vitanov_ARPC01,STIRAP_RMP98}, coherent \cite{BrumShap_CPL86,Shapiro_RMP07} and optimal control \cite{Tannor_JCP86,Kosloff_CP89}, to name a few. It is usually supposed that before the interaction with the laser field the system is in a particular state and the scheme is designed such that after the interaction it is found in another (the desired) state. In the case of two-level systems one talks about population inversion -- at the beginning the system is in one of the two states (usually the ground state) and after the manipulation it is completely transferred to the other one (usually an excited state). 

There are many interesting situations, however, when at the beginning the system is in a superposition of quantum states and we want to drive it to a new superposed state but controlling the weights with which each quantum state participates in the mixing. Therefore, we need a scheme which brings the system from $|S1\rangle = a|1\rangle + b|2\rangle$ to $|S2\rangle = c|1\rangle + d|2\rangle$ controlling the populations, i.e. $|c|^2$ and $|d|^2$. Obviously, the population inversion, i.e. going from $|1\rangle$ to $|2\rangle$, is a special case of this more general scheme.

An interesting example for a situation in which such a control protocol will be very useful can be realized when molecules are exposed to ultrashort laser pulses. In this case, due to the broad bandwidth of the pulse, more than one electronic states can be coherently populated creating in that way an electronic wave packet and therefore triggering pure electron dynamics. In the case of ionization of the molecule, it has been shown \cite{CM_first} that due to the electron correlation the initially created hole charge can migrate throughout the molecule within just few femtoseconds. This phenomenon, know as charge migration, has since been intensively studied theoretically \cite{Breidbach03,Hennig05,dprop,Luennemann_JCP08,RelaxSat,dipep_paper} and currently is a subject of increasing interest also form experimentalists (see, e.g., Refs.~\cite{MolAttoRev_CPL13,NatPhot14_comm,FrMiMa_NatPhot14}). The ultimate goal for the scientific efforts is to find schemes to control the charge migration. Ideally, an ultrashort pulse triggers electron dynamics and a delayed pulse is used for controlling the migration process.

Although most of the known schemes or protocols are applicable also to systems being in a superposition of quantum states, they do not allow for control of the final state mixture. Being designed for achieving a population inversion, applied to two-level systems in a superposed state these schemes will bring the system to a new superposed state in which the weights with which each eigenstate participates are just swapped. In this paper we show how one can obtain an analytic expression for a resonant pulse which is able to control not only the final state superposition, but also the exact path of the transition.

%\section{Main equations}

Let us briefly review the general formalism for describing the electric dipole interaction between a two-level system and a classical monochromatic field. The Hamiltonian in this case has the following form
\begin{equation}\label{ham}
	H = \epsilon_1 |1\rangle\langle 1| + \epsilon_2 |2\rangle\langle 2| - \vec d\cdot\vec{E}(t),
\end{equation}
where $|1\rangle$, $|2\rangle$ and $\epsilon_1$, $\epsilon_2$ are the two eigenstates and eigenenergies, respectively, of the field-free Hamiltonian, $\vec d$ is the electric dipole operator given by
\begin{equation}\label{dipol}
	\vec d = \vec e_d \left( d |1\rangle\langle 2| + d^*|2\rangle\langle 1| \right),
\end{equation}
with $\vec e_d$ being a unit vector in the direction of the dipole and $d$ denoting the matrix element of the dipole operator between $|1\rangle$ and $|2\rangle$. In the case of a laser pulse with carrier frequency $\omega$, the electric field  $\vec{E}(t)$ can be written as
\begin{equation}\label{field}
	\vec E(t) = \vec{\mathcal{E}}(t) e^{-i\omega t}+ \vec{\mathcal{E}}^*(t) e^{i\omega t},
\end{equation}
where $\vec{\mathcal{E}}(t)$ contains the polarization, amplitude, and envelope of the pulse. For simplicity, in the following we will take the dipole transition matrix element as real ($d=d^*$) and will remove the vector notations by taking the scalar product between $\vec d$ and $\vec E(t)$, denoting by $\mu$ the projection of the dipole operator on the polarization axis of the electric field.

The general form of the wave function describing the evolution of the system is given by (atomic units are used throughout the text unless otherwise specified)
\begin{equation}\label{wf}
	|\Psi(t)\rangle = c_1(t) e^{-i\epsilon_1 t}|1\rangle + c_2(t) e^{-i\epsilon_2 t} |2\rangle,
\end{equation}
where $c_1(t)$ and $c_2(t)$ are the time-dependent (in general complex) amplitudes of the eigenstates $|1\rangle$ and $|2\rangle$, which, due to the orthonormality of the field-free states satisfy the condition $|c_1(t)|^2+|c_2(t)|^2=1$ at all times.

Inserting this wave function in the time-dependent Schr\"odinger equation $i|\dot\Psi(t)\rangle=H|\Psi(t)\rangle$, one obtains the following set of coupled equations
\begin{subequations}\label{exact} 
\begin{eqnarray}
 \dot{c}_1(t) &=& ic_2(t)\mu\left(\mathcal{E}(t)e^{-i(\omega+\omega_0)t} + \mathcal{E}^*(t)e^{i(\omega-\omega_0)t}\right), \\
 \dot{c}_2(t) &=& ic_1(t)\mu\left(\mathcal{E}(t)e^{-i(\omega-\omega_0)t} + \mathcal{E}^*(t)e^{i(\omega+\omega_0)t}\right),
\end{eqnarray}
\end{subequations}
where $\omega_0 = \epsilon_2 - \epsilon_1$ denotes the resonance frequency. 

Since these equations are generally not solvable in analytical closed form, one usually introduces at this point the so-called rotating-wave approximation (RWA), meaning that the ``rapidly oscillating terms'', i.e. the exponentials $e^{\pm i(\omega+\omega_0)t}$ in Eqs.~(\ref{exact}), are neglected. This approximation usually works well if $\omega\approx\omega_0$ (near resonance) and the coupling to the field is not very strong. We note that we need the RWA only for obtaining an analytical expression for the control field. As we will see, this field will give the desired results also when the exact equations [Eq.~(\ref{exact})] for the system evolution are used. Within the RWA, Eqs.~(\ref{exact}) turn into the following set of equations
\begin{subequations}\label{rwa} 
\begin{eqnarray}
 \dot{c}_1(t) &=& i c_2(t)\mu\mathcal{E}^*(t)e^{i\delta t}, \\
 \dot{c}_2(t) &=& i c_1(t)\mu\mathcal{E}(t)e^{-i\delta t},
\end{eqnarray}
\end{subequations}
where $\delta=\omega-\omega_0$ denotes the detuning. Equations (\ref{rwa}) can be integrated exactly for arbitrary initial conditions to obtain a closed analytical solution. 

%\section{Reverse-engineering approach}

Our purpose is, however, to find pulses which will bring the system into a state with a desired proportion of the final populations, given by the modulus square of the amplitudes $c_1$ and $c_2$. In order to find such solutions we can view Eqs.~(\ref{rwa}) as equations for $\mathcal{E}(t)$ and $\mathcal{E}^*(t)$, i.e.
\begin{subequations}\label{fieldeqs} 
\begin{eqnarray}
 \mathcal{E}^*(t)&=&-\frac{i}{\mu} \frac{\dot{c}_1(t)}{c_2(t)} e^{-i\delta t}, \\
 \mathcal{E}(t)&=&-\frac{i}{\mu} \frac{\dot{c}_2(t)}{c_1(t)} e^{i\delta t}.
\end{eqnarray}
\end{subequations}

Substituting these expressions into Eq.~(\ref{field}) one obtains
\begin{equation}\label{pulse}
E(t)=-\frac{i}{\mu} \left(\frac{\dot{c}_2(t)}{c_1(t)} e^{-i\omega_0 t} +
	                  \frac{\dot{c}_1(t)}{c_2(t)} e^{i\omega_0 t} \right).	
\end{equation}

Note that due to the RWA the dependence on $\omega$ cancels out. Now we have an expression that connects the evolution of the amplitudes with the driving field. Therefore, if we want that the system evolves in a particular way, we can obtain through Eq.~(\ref{pulse}) the field which can drive this evolution.

As we mentioned, the amplitudes $c_1(t)$ and  $c_2(t)$ are in principle complex functions, so we can write them in the form
\begin{equation}\label{amplitudes}
 c_k(t)=\tilde c_k(t)e^{i\varphi_k}, \quad k=1,2,
\end{equation}
where $\tilde c_k(t)$ are real positive functions. In principle, $\varphi_k$ can be time-dependent. As we will see below, this would lead to a chirped pulse. Let us concentrate first on the solution for time-independent $\varphi_k$. In this case, Eq.~(\ref{pulse}) takes the form
\begin{equation}\label{pulse2}
E(t)=-\frac{i}{\mu} \left(\frac{\dot{\tilde{c}}_2(t)}{\tilde{c}_1(t)} e^{-i(\omega_0 t+\varphi)} +
	                  \frac{\dot{\tilde{c}}_1(t)}{\tilde{c}_2(t)} e^{i(\omega_0 t+\varphi)} \right),	
\end{equation}
where $\varphi=\varphi_1-\varphi_2$ is the relative phase between the amplitudes $c_1(t)$ and  $c_2(t)$.

Let the evolution of the system proceed according to the function $f(t)$, i.e. let $|\tilde c_1(t)|^2=f(t)$. From the total population conservation condition we automatically have $|\tilde c_2(t)|^2= 1-f(t)$. Since $\tilde c_k(t)$ are both real and positive, we can write that $\tilde c_1(t)=\sqrt{f(t)}$ and $\tilde c_2(t)=\sqrt{1-f(t)}$. From Eq.~(\ref{pulse2}) we get the following field 
\begin{equation}\label{controlf}
	E(t)=\frac{1}{\mu} \frac{\dot{f}(t)}{\sqrt{f(t)(1-f(t))}} \sin{(\omega_0 t + \varphi)},
\end{equation}
which has a carrier frequency exactly on resonance with the transition between the two states $\omega_0$.

We see that if we want to drive the system in a particular way, we just need to describe this evolution via an appropriate control function $f(t)$. The control field obtained via Eq.~(\ref{controlf}) will then ``force'' the system to follow the quantum path given by $f(t)$. We note that the control field derived in Eq.~(\ref{controlf}) will keep the relative phase between the amplitudes constant. 

If one allows for the phases in Eq.~(\ref{amplitudes}) to be time-dependent, an additional term would appear in Eq.~(\ref{pulse2}) containing the time derivatives of the individual phases. Since the field has to be real, one can, in principle, obtain conditions for an additional control function which can fix the way the relative phase evolves in time. However, the discussion of this more complicated construct is out of scope of the present paper, since we would like to concentrate here only on the problem of population control. 

%In principle, one can use a similar procedure assuming a time-dependent relative phase, i.e. taking $c_k(t)=\tilde c_k(t)e^{i\varphi_k(t)}$. This will lead to a chirped pulse ($\varphi$ in Eq.~(\ref{controlf}) will be time-dependent) and will introduce an additional term in the field containing the time derivatives of the individual phases. Since the field has to be real, one can, in principle, obtain conditions for an additional control function which can fix the way the relative phase evolves in time. However, in the present paper we would like to concentrate only on the problem of population control. % However, the discussion of this more complicated construct is out of scope of the present paper. 

Let us now illustrate the above reverse-engineering approach with a concrete analytical and numerical example. Suppose that initially our system is in a (superposed) state in which the population of $|1\rangle$ is equal to $a_i$ and we want to drive the system to a state in which the population of $|1\rangle$ will be $a_f$. A convenient choice for the function controlling this transition is
\begin{equation}\label{ex_f}
	f(t)=a_i(1-g(t))+a_f g(t),
\end{equation}
where $g(t)$ is a function which goes smoothly from 0 to 1 and never exceeds 1, i.e. $0 \le g(t) \le 1$, $g(t) \xrightarrow{t \to -\infty} 0$, and $g(t) \xrightarrow{t \to \infty} 1$. A possible choice for such a function is the following
\begin{equation}
	g(t)=\frac{1}{1+e^{-\alpha t}},
\end{equation}
where the parameter $\alpha$ controls the duration of the transition from $a_i$ to $a_f$.

If we use this particular choice of control function $f(t)$ in Eq.~(\ref{controlf}), we obtain the following control field
\begin{widetext}
 \begin{eqnarray}\label{example}
E(t)=\frac{1}{\mu} \frac{\alpha (a_f-a_i)e^{\alpha t}}{(1+e^{\alpha t})\sqrt{(1-a_i+(1-a_f)e^{\alpha t})(a_i+a_f e^{\alpha t})}} \sin{(\omega_0 t + \varphi)}.
\end{eqnarray}
\end{widetext}

The parameter $\alpha$ connects the speed of transition with the field intensity. For a faster transition we will naturally need to apply a stronger field. 

%\section{Example and numerical check}

Let us now perform a numerical test on this particular control field, taking the following numerical values for the parameters: $\mu=6$~au, $\omega_0=0.02$~au, $a_i=0.4$, $a_f=1$, $\alpha=0.01$, and $\varphi=0$. %Therefore, we have a two-level system which is initially in a non-stationary state, in which 40\% of the total population is in level $|1\rangle$ and the remaining 60\% are in $|2\rangle$, and we want to drive it such that we transfer the entire population into level $|1\rangle$. After the pulse, the system is in an eigenstate.

\begin{figure}[ht!]
\begin{center}
\includegraphics[width=8.5cm]{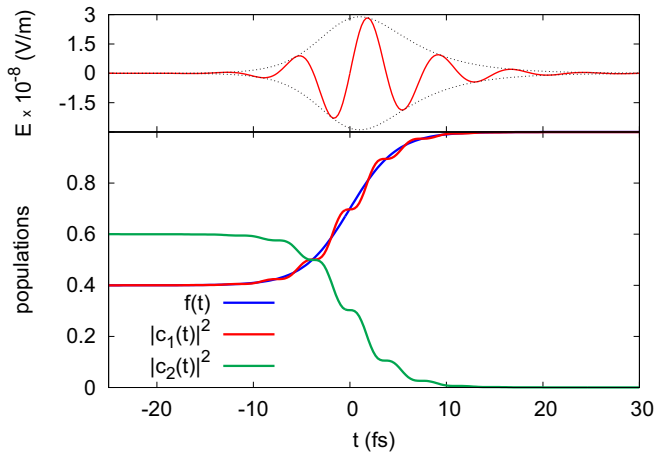}
\end{center}
\caption{\label{fig_pop1}(Color online) Upper panel: Laser pulse obtained through Eq.~(\ref{example}) using the following parameters: $\mu=6$~au, $\omega_0=0.02$~au, $a_i=0.4$, $a_f=1$, $\alpha=0.01$, and $\phi=0$. Lower panel: Evolution of the populations of the states $|1\rangle$ (red) and $|2\rangle$ (green) of a two-level system driven by the such a laser pulse. The control function $f(t)$ used to obtain the driving field is depicted in blue. We see that the system is driven from a coherent superposition of states $|1\rangle$ (40\%) and $|2\rangle$ (60\%) to a system being only in state $|1\rangle$.}
\end{figure}

The pulse corresponding to the above parameters is shown in the upper panel of Fig.~\ref{fig_pop1}. We see that it is slightly asymmetric with respect to the maximum intensity but otherwise very regular. The asymmetry actually reflects the ``asymmetric'' way we want to drive the system. The chosen parameters describe the situation in which the system is initially in a non-stationary state (a superposition of $|1\rangle$ and $|2\rangle$, with $|1\rangle$ containing 40\% of the total population) and we drive it to a stationary state, i.e. after the pulse the system is in state $|1\rangle$ (see the blue curve in lower panel of Fig.~\ref{fig_pop1}). The special case of population inversion, i.e. when the populations of the states $|1\rangle$ and $|2\rangle$ before and after the pulse are swapped, will require a symmetric pulse. If we take $a_f$ to be 0.6, Eq.~(\ref{example}) will give us a symmetric pulse. 

To check the validity of the above procedure, we can use the field obtained via Eq.~(\ref{example}) and solve numerically Eqs.~(\ref{exact}), i.e. the equations before introducing the rotating-wave approximation. We remind that within the dipole approximation for the interaction with the field, these are the exact equations describing the evolution of the system. The result is depicted in the lower panel of Fig.~\ref{fig_pop1}, together with the control function $f(t)$ giving the evolution of the system within the RWA. We see that the system indeed follows the desired evolution and that the RWA is a quite good approximation in this case. The neglected rapidly oscillating terms introduce only small variations in the transition path.

We note in passing that the parameters in this example are not taken arbitrarily. They describe the situation realized in the outer-valence ionization of the molecule MePeNNA \cite{Luennemann_JCP08}. Ultrafast ionization out of the highest occupied molecular orbital of MePeNNA coherently populates two ionic states, which are each composed of two configurations: charge located on the amine site (corresponding to state $|1\rangle$) and charge located on the chromophore site (corresponding to state $|2\rangle$). The coherent population of the two ionic states triggers charge-migration dynamics~-- the created hole-charge oscillates between the chromophore and the amine site of the molecule with a period of about 7.5~fs \cite{Luennemann_JCP08} (see also Ref.~\cite{emission}). Applying the laser pulse shown in the upper panel of Fig.~\ref{fig_pop1} will stop the quantum beating and localize the charge on the amine site. One can, of course, design a pulse which can terminate the oscillation and localize the charge on the chromophore, or create any desired superposition of the two states involved. We, therefore, have a tool to control the ultrafast charge-migration dynamics.

\begin{figure}[ht!]
\begin{center}
\includegraphics[width=8.5cm]{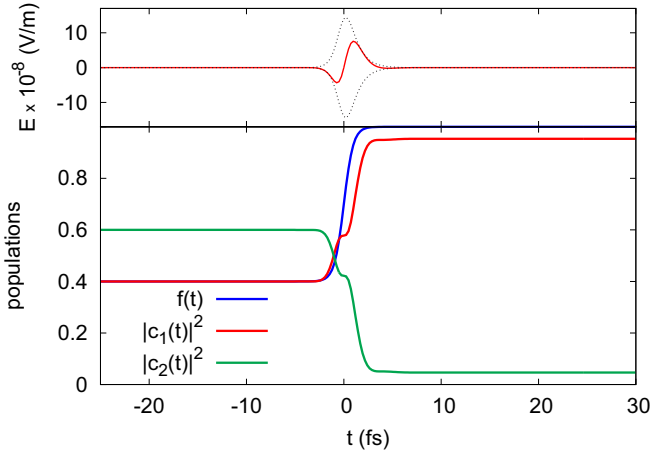}
\end{center}
\caption{\label{fig_pop2}(Color online) Upper panel: Laser pulse obtained through Eq.~(\ref{example}) using the same parameters as in Fig.~\ref{fig_pop1} except $\alpha=0.05$. Lower panel: Evolution of the populations of the states $|1\rangle$ (red) and $|2\rangle$ (green) of a two-level system driven by the such a laser pulse. The control function $f(t)$ used to obtain the driving field is depicted in blue. We see that the full solution start to deviate from the RWA and the pulse is not able to bring the system 100\% to state $|1\rangle$.}
\end{figure}

It is important to know what is the regime in which the above approach for obtaining the driving field works well, i.e. when the RWA is a good approximation. The condition for the validity of the RWA is that the envelop of the pulse varies slowly with time in comparison to the field oscillations, determined by the carrier frequency $\omega_0$. To exemplify this let us take the same parameters as above but reduce the time for which we want to drive the system to the chosen final state. The transition time is governed by the parameter $\alpha$ which also determines the width of the pulse envelope. The result for the pulse obtained with $\alpha=0.05$ (all other parameters were kept the same) is shown in Fig.~\ref{fig_pop2} together with the evolution of the system. With these parameters we obtain a single-cycle pulse with which, as we see in the lower panel of Fig.~\ref{fig_pop2}, the full solution starts to deviate more substantially from the one obtained within the RWA and we are no longer able to bring the system 100\% to state $|1\rangle$. However, even for such a limiting pulse the RWA still gives reasonable results.

At the end we would like to note that although focused on laser-driven population control, the scheme presented in the present paper is general and could be applied for designing other types of control fields. For example, systems often studied for the purposes of quantum computing are spin systems interacting with a magnetic field. In this case, the Hamiltonian has the same form as in Eq.~(\ref{ham}), just the interaction term has to be replaced by $-\vec\mu\cdot\vec{B}$, where $\vec\mu$ stands here for the magnetic moment and $\vec B$ for the magnetic field. As far as the resulting dynamical problem is the same, one may use the reverse-engineering approach presented here for obtaining magnetic fields that can drive a spin system in a desired way. The possibility to chose the way the system is driven from its initial to its final state is the main (and very important) difference to the scheme proposed in the present paper and other protocols for optimal control (see, e.g., Refs.~\cite{Sugny_PRA13,Sugny_PRL13}). In these schemes the control fields are obtained either by minimizing the transition time, or by minimizing the area of the pulse. In this respect, the present procedure is more general.

In summary, we proposed a simple method allowing to obtain laser pulses that can drive a two-level system in a desired way. Importantly, not only the final populations can be controlled. By choosing the function $f(t)$, we can exactly predetermine the evolution of the system and control the population of each state at any moment of time during the interaction with the field. We exemplified this on a system being in a superposition of the two states and showed that the analytically obtained resonant laser pulse can smoothly drive the system such that it lends on only one of the sates. This protocol may find an interesting application in controlling the ultrafast charge-migration process in molecules, but it is not restricted to systems interacting with laser fields. 

The authors thank Lorenz Cederbaum, Philipp Demekhin, and Shachar Klaiman for many valuable discussions. Financial support by the DFG is gratefully acknowledged.

\end{document}